%% file: menu07.tex
\documentclass[12pt,a4paper]{conference}

\usepackage{fancyhdr}
\usepackage{graphicx,amsmath,amssymb,cite}
\usepackage{multind}

\makeindex{author} \makeindex{subject}

\input MENU07macros.tex

\begin{document}

\begin{titlepage}
\begin{flushright}
LU TP 07-34\\
October 2007
\end{flushright}

\vfill

\begin{center}

{\large\bf Eta and Eta'
physics\footnote{Plenary talk given at MENU07, September 10-14,
2007 J\"ulich, Germany}}\\[2cm]

{\bf Johan Bijnens}\\[0.5cm]

Department of Theoretical Physics\\
Lund University\\
S\"olvegatan 14A\\
SE 22362 Lund, Sweden
\end{center}

\vfill

\begin{center}
\textbf{Abstract}
\end{center}
This talk describes the reasons why $\eta$ and $\eta^\prime$ decays are
an interesting topic of study for both theory and experiment. The main
part discusses the results of the recent calculation of $\eta\to3\pi$
at two-loop order in ChPT. Some puzzling aspects of the results
compared to earlier dispersive calculations are highlighted.
I also like to remind the reader of the use of $\eta$ and $\eta^\prime$ decays
for studying the anomaly.

\vfill
\end{titlepage}
\setcounter{page}{1}


\Chapter{ETA AND ETA' PHYSICS}
           {$\eta$ and $\eta'$ physics}{Johan Bijnens}
\vspace{-6 cm}\includegraphics[width=6 cm]{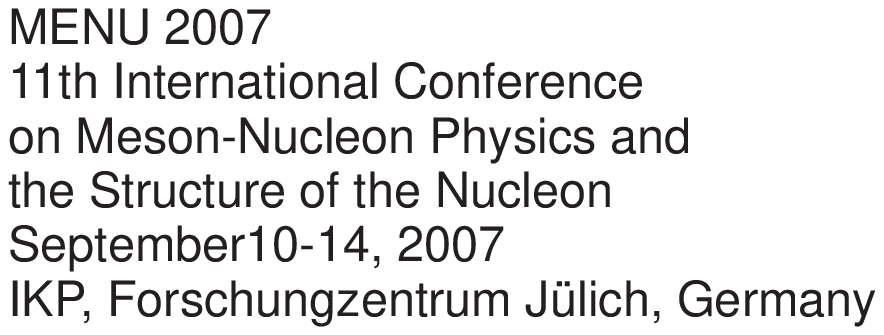}
\vspace{4 cm}

\addcontentsline{toc}{chapter}{{\it J. Bijnens}} \label{authorStart}

\begin{raggedright}

{\it Johan Bijnens
}
\index{author}{Bijnens, J.}\\
Department of Theoretical Physics\\
Lund University\\
S\"olvegatan 14A\\
SE 22362 Lund, Sweden
\bigskip\bigskip

\end{raggedright}

\begin{center}
\textbf{Abstract}
\end{center}
This talk describes the reasons why $\eta$ and $\eta^\prime$ decays are
an interesting topic of study for both theory and experiment. The main
part discusses the results of the recent calculation of $\eta\to3\pi$
at two-loop order in ChPT. Some puzzling aspects of the results
compared to earlier dispersive calculations are highlighted.
I also like to remind the reader of the use of $\eta$ and $\eta^\prime$ decays
for studying the anomaly.

\newcommand{\bee}{\begin{equation}}
\newcommand{\eee}{\end{equation}}
\newcommand{\ba}{\begin{eqnarray}}
\newcommand{\ea}{\end{eqnarray}}

\section{Introduction}

This conference has a lot of talks related to $\eta$ and $\eta^\prime$,
\index{subject}{eta decay}
both on decays, production and in a hadronic medium. The production is treated
in a plenary talk by Krusche and decays experimentally in the talk by
Wolke. There were also a lot of talks for both production and decay
in the parallel sessions. In this talk I will concentrate on decays
and in particular mainly on $\eta\to3\pi$. There are lots of references
treating $\eta$ and $\eta'$ physics. Many of them can be found in the
proceedings of two recent conferences devoted to them \cite{etahandbook,ETA05}.
There have also been more recent workshops in J\"ulich (ETA06) and
Peniscola (ETA07).

This talk first discusses why $\eta$ and $\eta^\prime$ are interesting, then
reminds the reader of some of the aspects of Chiral Perturbation Theory (ChPT)
after which the main part, devoted to $\eta\to3\pi$ comes. I close by
pointing out some properties of $\eta^\prime\to\eta\pi\pi,\pi\pi\pi$ decays
and the anomaly. Earlier reviews covering similar material are
Refs.~\cite{BG,BijnensETA05}.

\section{Why are $\eta$ and $\eta^\prime$ Interesting?}

The $\eta$ and $\eta^\prime$ are particles that decay strongly but all their
decays
are suppressed. That means that they are good laboratories to study non-dominant
strong interaction effects. Weak decays can happen but do occur at branching
ratios of order $10^{-11}$ or lower. So, if charge conjugation violation
would be discovered it would be very important. On the other hand, most
standard extensions of the standard model do not predict such effects at an
observable level in $\eta$ or $\eta^\prime$ decays.

But let us first see why pseudo-scalars are special. The QCD Lagrangian is
\bee
\mathcal{L}_{QCD} =  \sum_{q=u,d,s}
\left[i \bar q_L D\hskip-1.3ex/\, q_L +i \bar q_R D\hskip-1.3ex/\, q_R
- m_q\left(\bar q_R q_L + \bar q_L q_R \right)
\right]
\eee
So if $m_q = 0$ then the left and right handed quarks are decoupled
and they can be interchanged freely among themselves leading to a global
symmetry $G=U(3)_L \times U(3)_R$.
This symmetry is clearly broken in the hadron spectrum, the proton and the
$S_{11}$, as well as the $\rho$ and the $a_1$ have very different
masses\footnote{There is a discussion at present whether chiral
symmetry is restored for higher hadron masses. This is not relevant
for this talk. Recent references can be traced from \cite{Glozman,SV}.}.
The chiral symmetry group $G$ must thus be spontaneously broken, only the
vector part of the group is clearly visible in the spectrum.

As a consequence there must be a set of light particles, the pseudo-Goldstone
boson, whose interactions vanish at zero momentum as follows from Goldstone's
theorem. There are eight fairly light particles around with the right quantum
number, $\pi^0$, $\pi^\pm$, $K^\pm$, $K^0$, $\overline{K^0}$ and $\eta$.
But the next candidate with the correct quantum numbers, the $\eta^\prime$, is
heavy. We write the group $G$ in terms of simple groups,
\bee
G = U(3)_L\times U(3)_R = SU(3)_L\times SU(3)_R\times
 U(1)_V\times U(1)_A\,,
\eee
and notice that
the breaking pattern of $G=SU(3)_L\times SU(3)_R\longrightarrow H=SU(3)_V$
gives eight light particles as observed. The reason is that the $U(1)_A$
part of $G$ is a good symmetry of the classical action but not of
the full quantum theory. The divergence of its current
has a part coming from the anomaly
which couples to gluons via
\bee
\partial_\mu A^{0\mu} = 2 \sqrt{N_f}\omega\,
\quad\mbox{with}
\quad
\omega = \frac{1}{16\pi^2} \varepsilon^{\mu\nu\alpha\beta}
\,\mbox{tr}\,G_{\mu\nu}G_{\alpha\beta}\,.
\label{breakU1a}
\eee
$\omega$ is strongly interacting thus the divergence of the singlet
axial-current cannot be treated as zero. So the $\eta^\prime$ can be heavy as
is seen experimentally. Quantum effects break thus the $U(1)_A$, however
the r.h.s.\ of (\ref{breakU1a}) is a total divergence, so how can it have an
effect? The answer was found by `t~Hooft \cite{tHooft}. Gauge field
configurations with non-zero winding number, instantons, can produce an effect.
This in turn led to the so-called strong CP problem \emph{but} solved the
$\eta^\prime$ mass problem. A conclusion is thus that the $\eta^\prime$ has
potentially large and very interesting non-perturbative effects and
interactions with gluonic degrees of freedom that differ from other hadrons.
Since $\hat m\neq m_s$ this also affects $\eta$ physics via mixing.

\section{Chiral Perturbation Theory}

\index{subject}{Chiral perturbation theory}
The chiral symmetry of QCD and its spontaneous breaking has many consequences.
The best method to exploit these is Chiral Perturbation Theory (ChPT) which is
best defined via\\
\framebox{\parbox{0.985\textwidth}{ChPT $\equiv$
``Exploring the consequences of the chiral symmetry of QCD
and its spontaneous breaking using
effective field theory techniques.''}}
A derivation which clearly brings out all the assumptions involved
is \cite{Leutwyler0}. Lectures and review articles can be found in my Lattice07
talk \cite{lattice07} or on the webpage \cite{webpage}.
The original modern references are \cite{Weinberg0,GL0}.

ChPT uses as power-counting essentially dimensional counting in terms of a
generic momentum $p$. Momenta and
meson masses are counted as order $p$. Because of the Gell-Mann-Oakes-Renner
relation, $m_M^2 \propto m_q$, quark masses and external
scalar and pseudo-scalar
fields are counted as order $p^2$ and the covariant derivative requires external
vector and axial-vector field to be counted as order $p$.
With this counting there is no term of order $p^0$ in the chiral Lagrangian.
The lowest order Lagrangian is given by
\bee
{\cal L}_2 = \frac{F_0^2}{4} \{\langle D_\mu U^\dagger D^\mu U \rangle 
+\langle \chi^\dagger U+\chi U^\dagger \rangle \}\, ,
\eee
with $U$
parameterizing the Goldstone Boson manifold $G/H$ with
\bee
U(\phi) = \exp(i \sqrt{2} \Phi/F_0)\,,
\quad
\Phi (x) = \, \left( \begin{array}{ccc}
\displaystyle\frac{ \pi^0}{ \sqrt 2} \, + \, \frac{ \eta_8}{ \sqrt 6}
 & \pi^+ & K^+ \\
\pi^- &\displaystyle - \frac{\pi^0}{\sqrt 2} \, + \, \frac{ \eta_8}
{\sqrt 6}    & K^0 \\
K^- & \bar K^0 &\displaystyle - \frac{ 2 \, \eta_8}{\sqrt 6}
\end{array}  \right) .
\eee
The external fields are in the covariant derivative,
$
D_\mu U = \partial_\mu U -i r_\mu U + i U l_\mu
$
for the left and right external currents: $
r(l)_\mu = v_\mu +(-) a_\mu$ and the external scalar
and pseudo-scalar external densities are in $\chi = 2 B_0 (s+ip)$.
Quark masses come via the scalar density
 $s= {\cal M} + \cdots
$ and traces are over (quark) flavours
$\langle A \rangle = Tr_F\left(A\right)$.
The number of parameters increases fast at higher orders, there are
10+2 at order $p^4$\cite{GL1} and 90+4 at order $p^6$ \cite{BCE1}
for three-flavour mesonic ChPT.

\begin{figure}
\begin{center}
\includegraphics[width=0.9\textwidth]{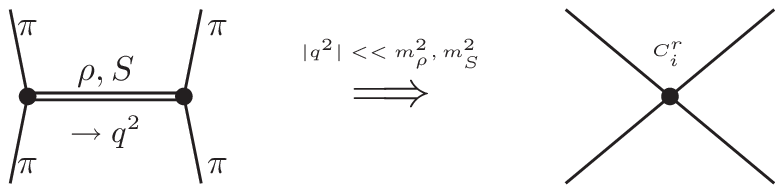}
\end{center}
\caption{\label{figCi}
Resonance saturation of the order $p^6$ low-energy-constants
$C_i^r$ via resonate exchange for $\pi\pi$-scattering.}
\end{figure}

The main uses of ChPT are that it contains all the $SU(3)_V$ relations
automatically and in addition relates processes with different numbers of
pseudo-scalars and it includes the nonanalytic dependences on masses and
kinematical quantities, often referred to as chiral logarithms. As an example,
the pion mass in two-flavour ChPT is given by \cite{GL0}
\bee
 m_\pi^2 = 2 B \hat m  + \left(\frac{2 B \hat m}{F}\right)^2
\left[ \frac{1}{32\pi^2}{ \log\frac{\left(2 B \hat m\right)}{\mu^2}} +
 2 l_3^r(\mu)\right] +\cdots\,,
\label{pionmass}
\eee
with $M^2 = 2 B \hat m$ and $B\neq B_0$, $F\neq F_0$ because of
two versus three-flavour ChPT. In (\ref{pionmass}) we see the logarithm and
the occurrence of the higher order parameter $\ell_3^r(\mu)$.
Eq.~(\ref{pionmass}) also shows some of the choices that need to be made
when performing higher order ChPT calculations: Which subtraction scale $\mu$
and which quantities should be used to express the results. Lowest order
masses or physical meson masses and dito for the decay constants and
other kinematical quantities as $s,t,u$ in $\pi\pi$-scattering. There is
clearly no unique choice and the choice can influence the \emph{apparent}
convergence of the ChPT series quite strongly. Another problem is that
typically, not all the higher
order parameters that show up in the calculations are known experimentally.
Thus one needs to make estimates of these, usually via
a version of resonance saturation originally introduced
in ChPT in \cite{EGPR}. This is schematically depicted in Fig.~\ref{figCi}.
More recent references on resonance saturation and possible pitfalls
are \cite{Cirigliano1,BGLP}. Discussions on this problem can also be
found in the papers on order $p^6$ ChPT and the review \cite{reviewp6}.

\section{$\eta\to3\pi$}

\index{subject}{eta decay to three pions}
In the limit of conserved isospin, i.e.\ we turn off electromagnetism and
set $m_u=m_d$, the $\eta$ is stable. Direct electromagnetic effects have been
known to be small since long ago \cite{Sutherland1,Sutherland2}.
It should thus proceed mainly through the quark-mass difference $m_u-m_d$.
The lowest order was done in \cite{orderp2x1,orderp2x2},
order $p^4$ in
\cite{GL3} and recently the full order $p^6$ has been evaluated \cite{BG07}.
In this section I will mainly present the new results of \cite{BG07}.

\begin{figure}
\begin{minipage}{0.48\textwidth}
\includegraphics[width=\textwidth]{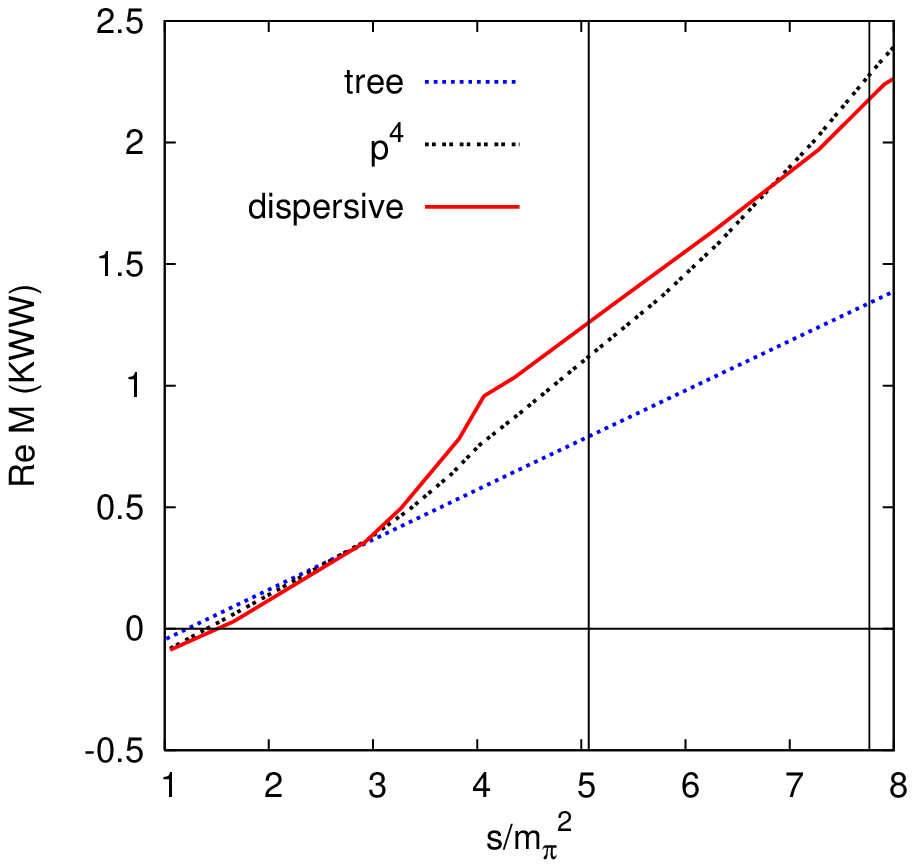}
\centerline{(a)}
\end{minipage}
\begin{minipage}{0.48\textwidth}
\includegraphics[width=\textwidth]{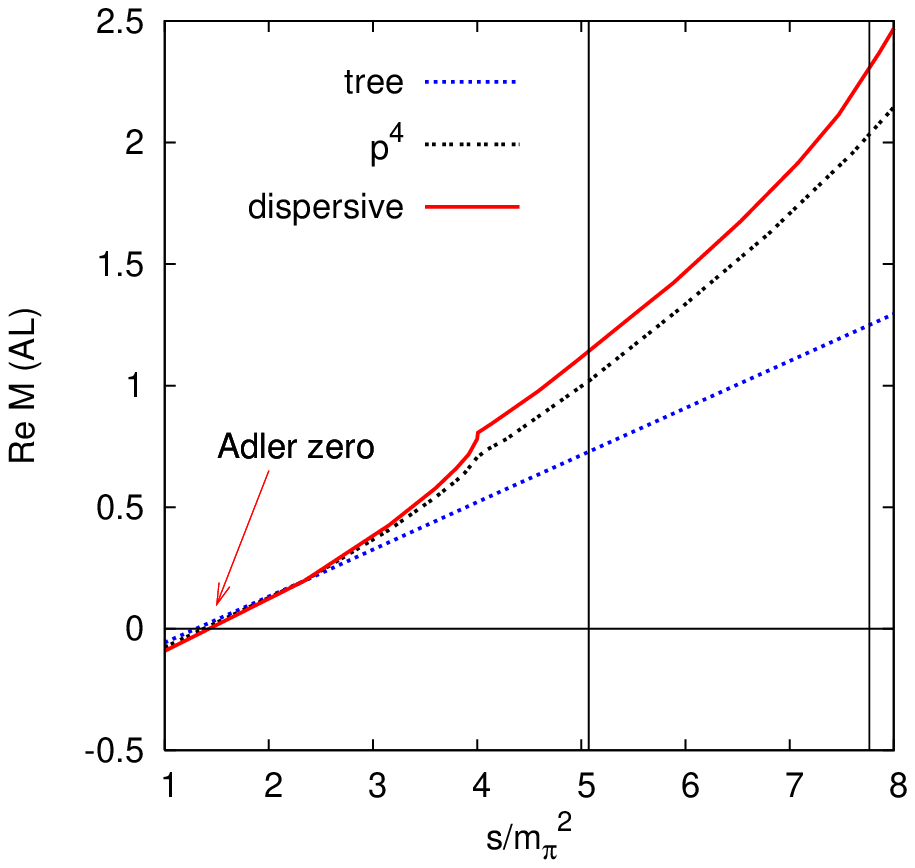}
\centerline{(b)}
\end{minipage}
\caption{
(a) Decay amplitude obtained by use of extended Khuri-Treiman
equations\cite{KWW} along the line $s=u$.
(b) Alternative dispersive analysis for the decay amplitude\cite{AL}.
Figs. from \cite{BijnensETA05}, adapted from \cite{KWW,AL}.
}
\label{figdispersive}
\end{figure}

The momenta for the decay $\eta\to\pi^+\pi^-\pi^0$ are labeled as
$p_\eta$, $p_+$, $p_-$ and $p_0$ respectively and we introduce the
kinematical Mandelstam variables
\bee
s = (p_+ +p_-)^2\,, 
t = (p_+ +p_0)^2\,, 
\nonumber\\
u = (p_- +p_0)^2\,. 
\label{defstu}
\eee
These are linearly dependent,
$
s+t+u = m_{\pi^{o}}^2 + m_{\pi^{-}}^2 + m_{\pi^{+}}^2 + m_{\eta}^2
\equiv 3 s_0\,.  
$
The amplitude is for the charged and neutral decay
\ba
\langle \pi^0\pi^+\pi^-{\mbox {out}}|\eta\rangle &=& i\left(2\pi\right)^4 
\,\delta^4\left(p_\eta-p_{\pi^+}-p_{\pi^-}-p_{\pi^0}\right)
\,A(s,t,u)\,,
\nonumber\\
\langle \pi^0\pi^0\pi^0{\mbox {out}}|\eta\rangle &=& i\left(2\pi\right)^4 
\,\delta^4\left(p_\eta-p_{1}-p_{2}-p_{3}\right)
\,\overline{A}(s_1,s_2,s_3)\,,
\nonumber\\
\overline{A}(s_1,s_2,s_3) &=& A(s_1,s_2,s_3)+A(s_2,s_3,s_1)+A(s_3,s_1,s_2)\, .
\label{defamplitude}
\ea
The relation in the last line of (\ref{defamplitude}) is only valid
to first order in $m_u-m_d$. The factor of $m_u-m_d$ can be pulled out in
various ways. Two common ones are
\bee
A(s,t,u) = \frac{\sqrt{3}}{4R}M(s,t,u)
\quad\mbox{or}
\quad
A(s,t,u) =\frac{1}{Q^2} \frac{m_K^2}{m_\pi^2}(m_\pi^2-m_K^2)\,
\frac{ { {\cal M}(s,t,u)}}{3\sqrt{3}F_\pi^2}\,,
\label{defM}
\eee
with quark-mass ratios $R= (m_s-\hat m)/(m_d-m_u)$
and $Q^2 = R(m_s+m_d)/(2\hat m)$.
The lowest order result corresponds to
\bee
M(s,t,u)_{LO} = \left(({4}/{3})\,m_\pi^2-s\right)/F_\pi^2\,.
\label{LO}
\eee
The tree level determination of $R$ in terms of meson masses gives
with (\ref{LO}) a decay rate of 66~eV which should be compared with
the experimental results of 295$\pm$17~eV\cite{PDG06}.
In principle, since the decay rate is proportional to $1/R^2$ or $1/Q^4$,
this should allow for a precise determination of $R$ and $Q$. However,
the change required seems somewhat large. The order $p^4$ calculation
\cite{GL3} increased the predicted decay rate to 150~eV albeit with a
large error. About half of the enhancement in the amplitude came from
$\pi\pi$ rescattering and the other half from other effects like the
chiral logarithms\cite{GL3}. The rescattering effects have been
studied at higher orders using dispersive methods in \cite{KWW}
and \cite{AL}. Both calculations found a similar enhancement in the decay rate
bringing it to about 220~eV but differ in the way the Dalitz plot distributions
look. This can be seen in Fig.~\ref{figdispersive} where I show the 
real part of the amplitude as a function of $s$ along the line $s=u$.
The calculations use a very
different formalism but make similar approximations, they mainly differ in
the way the subtraction constants are determined.
That discrepancy and the facts that in $K_{\ell4}$ the dispersive estimate
\cite{BCG}
was about half the full ChPT calculation \cite{ABT3} and
at order $p^4$ the dispersive effect was about half of the correction for
$\eta\to3\pi$ makes it clear that also for this process a full order $p^6$
calculation is desirable. This has been done recently in \cite{BG07}.

\begin{figure}
\begin{minipage}{0.515\textwidth}
\includegraphics[width=\textwidth]{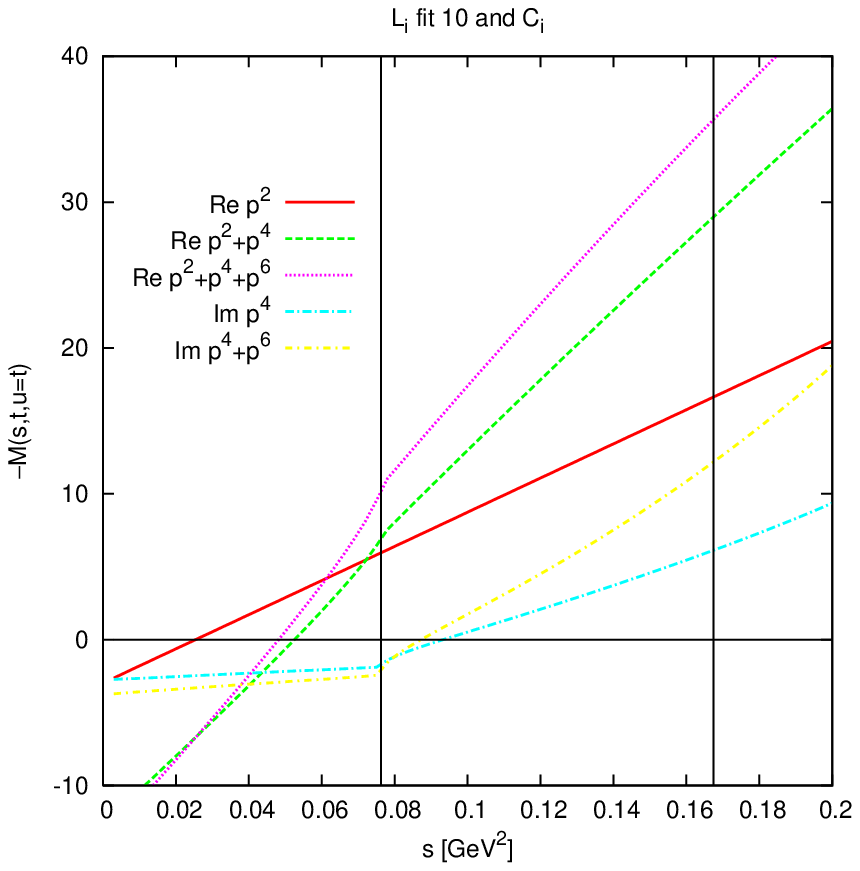}
\centerline{(a)}
\end{minipage}
\begin{minipage}{0.48\textwidth}
\includegraphics[width=\textwidth]{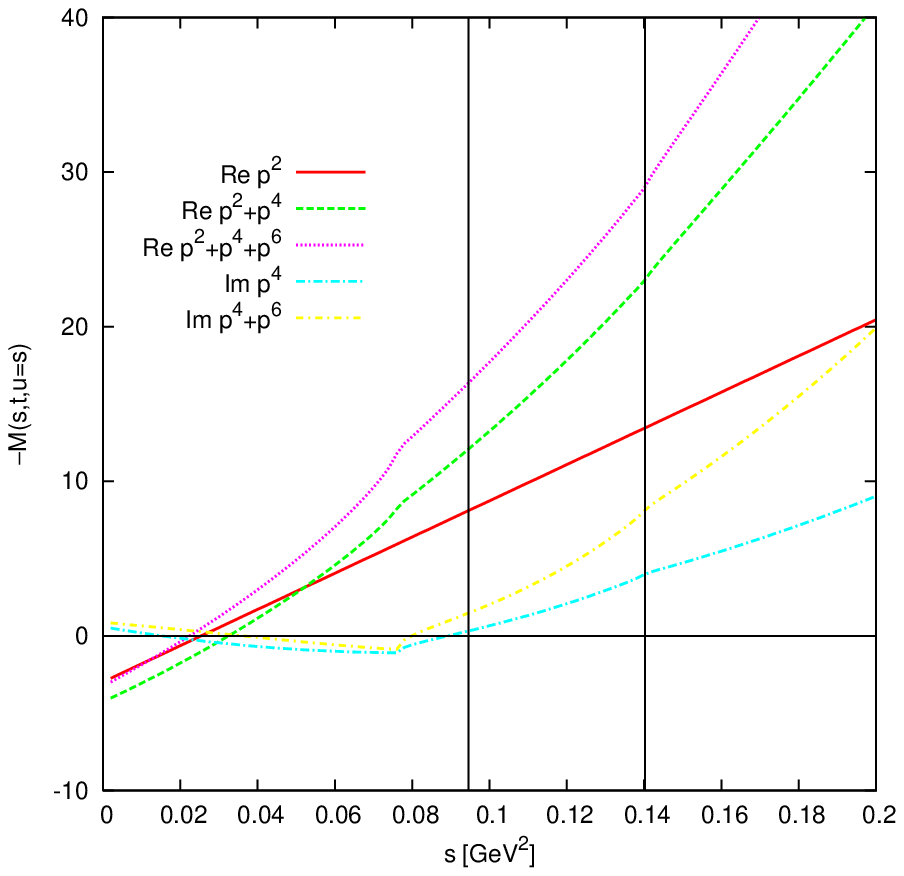}
\\[2mm]
\centerline{(b)}
\end{minipage}
\caption{(a) The amplitude $M(s,t,u)$ along the line $t=u$. The vertical
lines indicate the physical region.
Shown are the
real and imaginary parts with all parts summed up to the given order.
(b) Similar plot but along the line $s=u$.
Figs. from \cite{BG07}.}
\label{figMstu}
\end{figure}

Ref.~\cite{BG07} generalizes the methods
of \cite{ABT4} to deal with $\pi^0$-$\eta$ mixing to processes with mixing
on more than one external leg. The input parameters are from the main order
$p^6$ fit, called fit 10, of \cite{ABT4} and the needed order $p^6$ constants
are determined by resonance exchange as discussed earlier. Details can be
found in \cite{BG07}. In Fig.~\ref{figMstu} I show the numerical result
for the amplitude along two lines in the Dalitz plot, $t=u$ and $s=u$.
The latter can be compared directly with the dispersive result of
Fig.~\ref{figdispersive}.
The correction found in \cite{BG07} at order $p^6$ is 20-30\% in amplitude,
larger in magnitude than the dispersive estimates \cite{KWW,AL} but
with a shape similar to \cite{AL}.

The Dalitz plot in $\eta\to3\pi$ is parameterized in terms of $x$ and $y$
defined in terms of the kinetic energies of the pions $T_i$ and
$Q_\eta=m_\eta-2m_{\pi^+}-m_{\pi^0}$ for the charged decay and $z$ defined in
terms of the pion energies $E_i$. The amplitudes are expanded in $x,y,z$.
\ba
x &=& \sqrt3 \frac{T_+-T_-}{Q_\eta}\,,\quad y= \frac{3T_0}{Q_\eta}-1\,,\quad
z = \frac{2}{3}\sum_{i=1,3}\left(\frac{3 E_i-m\eta}{m_\eta-3m_{\pi^0}}\right)^2
\,,
\nonumber\\
|M(s,t,u)|^2 &=& A_0^2\left(1+ay+by^2+dx^2+fy^3+\cdots\right)\,,
\nonumber\\
|\overline M(s,t,u)|^2 &=& \overline A_0^2
\left(1+2\alpha_2+\cdots\right)\,.
\ea
Recent experimental results for these parameters are shown in 
Tabs.~\ref{tabDalitzcharged} and \ref{tabDalitzneutral}.
There are discrepancies among the experiments but the two latest
precision experimental
measurements of $\alpha$ agree.
\begin{table}
\caption{Measurements of the Dalitz plot distributions in 
$\eta\to\pi^+\pi^-\pi^0$. 
The KLOE result \cite{KLOEcharged} for $f$ is $f=0.14\pm0.01\pm0.02$.}
\label{tabDalitzcharged}
\begin{tabular}{|c|ccc|}
\hline
Exp. & a & b & d  \\
\hline
\rule{0cm}{12pt}KLOE \cite{KLOEcharged} & $-1.090$\small$\pm0.005^{+0.008}_{-0.019}$ &
 $0.124$\small$\pm0.006\pm0.010$ & $0.057$\small$\pm0.006^{+0.007}_{-0.016}$\\
CB \cite{CBarrelcharged} & $-1.22\pm0.07$ & $0.22\pm0.11$ &
$0.06$\small$\pm0.04$ (input) \\
 \cite{Layter73} & $-1.08\pm0.014$ & $0.034\pm0.027$ &
$0.046\pm0.031$ \\
\cite{Gormley} &$-1.17\pm0.02$ & $0.21\pm0.03$ 
& $0.06\pm0.04$ \\
\hline
\end{tabular}
\end{table}
\index{subject}{Dalitz plot distributions, eta decay}
The predictions from ChPT to order $p^6$ with the input parameters fixed
as described earlier are give in Tabs.~\ref{tabDalitzcharged_theory}
and \ref{tabDalitzneutral_theory}. The predictions from the dispersive
analysis as well as \cite{Borasoy} have not been included.
The different lines corresponds to variations on the input and the order of
ChPT. The lines labeled NNLO are the central results.
The agreement with experiment is not too good and clearly needs further study.
Especially puzzling is the $\alpha$ is consistently positive while
the dispersive calculations as well as \cite{Borasoy}
give a negative value.
The inequality $\alpha\le\left(d+b-a^2/4\right)/4$ derived in \cite{BG07} shows that
$\alpha$ has rather large cancellations inherent in its prediction and that
the overestimate of $b$ is a likely cause of the wrong sign obtained for
$\alpha$. In addition, the fairly large correction obtained gives in the end
somewhat larger values of $Q$ compared to those derived from the masses
\cite{BG07}.
\begin{table}
\caption{Measurements of the Dalitz plot distribution in
$\eta\to\pi^0\pi^0\pi^0$.}
\label{tabDalitzneutral}
\centerline{\begin{tabular}{|c|c|}
\hline
Exp. & $\alpha$\\
\hline
\rule{0cm}{12pt}
KLOE \cite{KLOEneutral2}      & $-0.027\pm0.004^{+0.004}_{-0.006}$\\
Crystal Ball \cite{CBall} & $-0.031\pm0.004$ \\
WASA/CELSIUS \cite{WASAneutral} & $-0.026\pm0.010\pm0.010$ \\
\hline
\end{tabular}}
\end{table}
\begin{table}
\caption{Theoretical estimate of the Dalitz plot distributions in 
$\eta\to\pi^+\pi^-\pi^0$.}
\label{tabDalitzcharged_theory}
\centerline{\begin{tabular}{|c|ccccc|}
\hline
               & $A_0^2$ & a & b & d & f \\
\hline
LO             & 120 & $-1.039$ & $0.270$ & $0.000$   & $0.000$ \\
NLO             & 314 & $-1.371$  & $0.452$ & $0.053$ & $0.027$\\
NLO ($L_i^r=0$) & 235 & $-1.263$& $0.407$ & $0.050$ & $0.015$\\
NNLO           & 538 &   $-1.271$ & $0.394$ & $0.055$ & $0.025$ \\
NNLO ($\mu=0.6$~GeV) & 543 & $-1.300$ & $0.415$ & $0.055$ & $0.024$\\         
NNLO ($\mu=0.9$~GeV) & 548 & $-1.241$ & $0.374$ & $0.054$ & $0.025$\\         
NNLO ($C_i^r = 0$)& 465 & $-1.297$  & $0.404 $ & $0.058$ & $0.032$ \\
NNLO ($L_i^r=C_i^r = 0$)& 251 & $-1.241$  & $0.424$ & $0.050$ & $0.007$ \\
\hline
\end{tabular}}
\end{table}
\begin{table}
\caption{Theoretical estimates of the Dalitz plot distribution in
$\eta\to\pi^0\pi^0\pi^0$. }
\label{tabDalitzneutral_theory}
\centerline{\begin{tabular}{|c|cc|}
\hline
    & $\overline A_0^2$ & $\alpha$\\
\hline
LO &  1090 & $ 0.000 $ \\
NLO & 2810 & $0.013$\\
NLO ($L_i^r=0$) & 2100 &   $ 0.016 $ \\
NNLO & 4790 & $ 0.013$ \\
NNLO ($C_i^r = 0$) & 4140 & $ 0.011$\\
NNLO ($L_i^r=C_i^r = 0$) & 2220 & $ 0.016$\\
\hline
\end{tabular}}
\end{table}

\section{Other Remarks}

I would simply like to repeat here some remarks made earlier,
see e.g. \cite{BijnensETA05}. The hadronic decays of the $\eta^\prime$
are interesting, they are predicted to be small at lowest
order. $\eta^\prime\to3\pi$ agrees reasonably well with expectations but
$\eta^\prime\to\eta\pi\pi$ has very large higher order corrections since
the lowest order is suppressed by a factor of $m_\pi^2$. I would also like
to emphasize once more that the decay of $\eta$ and $\eta^\prime$ allow
many tests of the triangle, quadrangle, $\ldots$ anomaly.

\section*{Acknowledgments}

This work is supported in part by the European Commission RTN network,
Contract MRTN-CT-2006-035482  (FLAVIAnet), 
the European Community-Research Infrastructure
Activity Contract RII3-CT-2004-506078 (HadronPhysics) and
the Swedish Research Council.
 



%
\end{document}

%% file: MENU07macros.tex
\newcommand{\beq}{\begin{equation}}
\newcommand{\eeq}[1]{\label{#1}\end{equation}}
\newcommand{\eeqn}{\end{equation}}

\newcommand{\beqa}{\begin{eqnarray}}
\newcommand{\eeqa}[1]{\label{#1}\end{eqnarray}}
\newcommand{\eeqan}{\end{eqnarray}}

\let\bar=\overbar

\newcommand{\Dslash}{\not{\hbox{\kern-4pt $D$}}}
\newcommand{\dslash}{\not{\hbox{\kern-2pt $\del$}}}

\newcommand{\msb}{{\bar{\ssstyle M \kern -1pt S}}}




%% file: menu07.bbl
\begin{thebibliography}{00}  


\bibitem{etahandbook}
  J.~Bijnens, G.~F\"aldt and B.M.K.~Nefkens (eds.)
  {\it Phys. Scripta}\ {\bf T99}, 1-282 (2002).

\bibitem{ETA05}
  B. H\"oistad and P. Moskal (eds.),
  {\it Acta Phys.\ Slov.}\  {\bf 56} 193-409 (2005).

\bibitem{BG}
  J.~Bijnens and J.~Gasser,
  {\it Phys.\ Scripta} {\bf T99}, 24 (2002)
  [hep-ph/0202242].

\bibitem{BijnensETA05}
  J.~Bijnens,
  {\it Acta Phys.\ Slov.}\  {\bf 56}, 305 (2005)
  [hep-ph/0511076].

\bibitem{Glozman}
  L.~Y.~Glozman,
  arXiv:0710.0978 [hep-ph].

\bibitem{SV}
  M.~Shifman and A.~Vainshtein,
  arXiv:0710.0863 [hep-ph].

\bibitem{tHooft}
  G.~`t Hooft,
  {\it Phys.\ Rev.}\   {\bf D14}, 3432 (1976)
  [Erratum\  {\bf D18}, 2199 (1978)].

\bibitem{Leutwyler0}
  H.~Leutwyler,
  {\it Ann. Phys.}\  {\bf 235}, 165 (1994)
  [hep-ph/9311274].

\bibitem{lattice07}
  J.~Bijnens,
  {\it PoS LATTICE 2007} (2007) 004
  [arXiv:0708.1377 [hep-lat]].

\bibitem{webpage}
  {\tt http://www.thep.lu.se/$\sim$bijnens/chpt.html}.

\bibitem{Weinberg0}
  S.~Weinberg,
  {\it Physica}\  {\bf A96}, 327 (1979).

\bibitem{GL0}
  J.~Gasser and H.~Leutwyler,
  {\it Annals Phys.}\  {\bf 158}, 142 (1984).

\bibitem{GL1}
  J.~Gasser and H.~Leutwyler,
  {\it Nucl.\ Phys.}\  {\bf B250}, 465 (1985).


\bibitem{BCE1}
  J.~Bijnens {\it et al.}, 
  {\it JHEP} {\bf 9902} (1999) 020
  [hep-ph/9902437].

\bibitem{EGPR}
  G.~Ecker {\it et al.}, 
  {\it Nucl.\ Phys.}\   {\bf B321} 311 (1989).

\bibitem{Cirigliano1}
  V.~Cirigliano {\it et al.}, 
  {\it Nucl.\ Phys.}\   {\bf B753}, 139 (2006)
  [hep-ph/0603205].

\bibitem{BGLP}
  J.~Bijnens {\it et al.}, 
  {\it JHEP} {\bf 0304} (2003) 055
  [hep-ph/0304222].

\bibitem{reviewp6}
  J.~Bijnens,
  {\it Prog.\ Part.\ Nucl.\ Phys.}\  {\bf 58}, 521 (2007)
  [hep-ph/0604043].

\bibitem{Sutherland1}
  D. G.~Sutherland
  {\it Phys. \ Lett.} {\bf 23}, 384 (1966).

\bibitem{Sutherland2}
  J.~S.~Bell and D.~G.~Sutherland,
  {\it Nucl.\ Phys.}\   {\bf B4}, 315 (1968).

\bibitem{orderp2x1}
  J.~A.~Cronin,
  {\it Phys.\ Rev.}\  {\bf 161}, 1483 (1967).

\bibitem{orderp2x2}
  H.~Osborn and D.~J.~Wallace,
  {\it Nucl.\ Phys.}\   {\bf B20}, 23 (1970).

\bibitem{GL3}
  J.~Gasser and H.~Leutwyler,
  {\it Nucl.\ Phys.}\  {\bf B250}, 539 (1985).

\bibitem{BG07}
  J.~Bijnens and K.~Ghorbani,
  arXiv:0709.0230 [hep-ph].


\bibitem{PDG06}
 W.~M.~Yao {\it et al.}  [Particle Data Group],
  {\it J.\ Phys.}\  {\bf G33}, 1 (2006).


\bibitem{KWW}
  J.~Kambor {\it et al.}, 
  {\it Nucl.\ Phys.}\   {\bf B465}, 215 (1996)
  [hep-ph/9509374].

\bibitem{AL}
    A.~V.~Anisovich and H.~Leutwyler,
    {\it Phys.\ Lett.}\  {\bf B375}, 335 (1996)
    [hep-ph/9601237].

\bibitem{BCG}
  J.~Bijnens {\it et al.}, 
  {\it Nucl.\ Phys.}\   {\bf B427}, 427 (1994)
  [hep-ph/9403390].

\bibitem{ABT3}
  G.~Amor\'os {\it et al.}, 
  {\it Nucl.\ Phys}.\   {\bf B585}, 293 (2000)
  [hep-ph/0003258].


\bibitem{ABT4}
  G.~Amor\'os {\it et al.}, 
  {\it Nucl.\ Phys.}\ {\bf B602}, 87 (2001)
  [hep-ph/0101127].

\bibitem{KLOEcharged}
 F.~Ambrosino {\it et al.}  
  arXiv:0707.2355 [hep-ex].

\bibitem{CBarrelcharged}
  A.~Abele {\it et al.}  
  {\it Phys.\ Lett.}\ {\bf B417}, 197 (1998).

\bibitem{Layter73}
  J.~G.~Layter {\it et al.}, 
  {\it Phys.\ Rev.}\  {\bf D7}, 2565 (1973).

\bibitem{Gormley}
  M.~Gormley{\it et al.}, 
  {\it Phys.\ Rev.}\ {\bf D2}, 501 (1970).

\bibitem{KLOEneutral2}
  F.~Ambrosino {\it et al.}  
  arXiv:0707.4137 [hep-ex].

\bibitem{CBall}
  W.~B.~Tippens {\it et al.}  
  {\it Phys.\ Rev.\ Lett.}\  {\bf 87}, 192001 (2001).

\bibitem{WASAneutral}
  M.~Bashkanov {\it et al.},
  arXiv:0708.2014 [nucl-ex].

\bibitem{Borasoy}
  B.~Borasoy and R.~Nissler,
  {\it Eur.\ Phys.\ J.} {\bf A26}, 383 (2005)
  [hep-ph/0510384].

\end{thebibliography}
